\documentstyle[aps,preprint,graphicx]{revtex}

\begin{document}
\title{Anomalous metallicity and electronic phase separation in the CsC$_{60}$
polymerized fulleride}
\author{Barbara Simovi\v{c}$^{(1)}$\cite{Auth1}, Denis
J\'{e}rome$^{(1)}$, and L\'{a}szl\'{o} Forr\'{o}$^{(2)}$}
\address{$^{(1)}$ Laboratoire de Physique des Solides (associ\'{e} au CNRS),
Universit\'{e} Paris-Sud, 91405 Orsay, France.\\$^{(2)}$
Laboratoire de Physique des Solides Semicristallins, EPFL,
Switzerland.}
\date{\today}
\maketitle

\begin{abstract}
$^{133}\mathrm{Cs}$ and $^{13}\mathrm{C}$-NMR have been used to
study the electronic properties of the polymerized phase of
$\mathrm{CsC}_{60}$ at ambient and under hydrostatic pressure. The
salient result of this study is the finding of fluctuations in the
local field at $^{133}\mathrm{Cs}$ site which are independent of
the applied pressure and due to thermally activated changes in the
local electronic environment of $^{133}\mathrm{Cs}$ nuclei. We
establish that the phase separation between magnetic and
nonmagnetic domains observed in the low temperature state at
ambient pressure is the result of a slowing down of these
fluctuations likely related to polaronic charge excitations on the
polymers.
\end{abstract}


\section{INTRODUCTION}
 When the face centered cubic (f.c.c) phase of ${\rm A}%
_{1}{\rm C}_{60}$(A=K, Rb, Cs) compounds is slowly cooled from
400K, one-dimensional polymerization of C$_{60}$ molecules
spontaneously occurs along the (110) cubic direction and leads to
an orthorhombic phase\cite {Stephens94}. A drastic change in the
electronic properties is observed at the structural
transition\cite{Tycko93,Chauvet94}. Indeed, the f.c.c. phase was
shown to be a Mott insulator\cite{Tycko93,Chauvet94}, whereas in
the orthorhombic phase a plasma frequency was measured in optical
experiments \cite{Bommeli95}. However, the density of carriers is
likely to be rather low or their effective mass very large since
the plasma frequency is equal to 0.1eV in ${\rm KC}_{60}$ polymer
and even lower for ${\rm RbC}_{60}$ and ${\rm
CsC}_{60}$\cite{Bommeli95}. In addition, the low frequency
conductivity of both ${\rm RbC}_{60}$ and ${\rm CsC}_{60}$
decreases
smoothly over a broad temperature range, being at variance with ${\rm KC}%
_{60}$ which remains conducting down to 4.2K\cite{Bommeli95}.
Furthermore, the temperature dependence of the $^{13}{\rm C}$
spin-lattice relaxation rate shows that strong magnetic
fluctuations are present up to room temperature in ${\rm
RbC}_{60}$ and ${\rm CsC}_{60}$\cite{Brouet96} and the sharp
decrease of the uniform static susceptibility (measured from EPR
line intensity) below 50K for ${\rm RbC}_{60}$ and 40K for ${\rm
CsC}_{60}$\cite{Bommeli95}, suggests that both compounds undergo
magnetic transitions at these respective temperatures. The
occurrence of spin ordering is also evident from NMR
experiments\cite {Brouet96,Senzier96}: the slowing down of
magnetic fluctuations gives rise to a divergent relaxation rate
below 40K. However, the nature of the spin order is less obvious.
On one hand, EPR experiments \cite{Janossy97,Bennati98} suggest
the onset of a spin density wave ground state as a result of a
possible one-dimensional (1D) character of the band structure. On
the other hand, $\mu {\rm SR}$ studies\cite{MacFarl95,Uemura95}
show a gradual transition towards a highly disordered magnetic
phase and do not rule out the possibility of a random spin
freezing below 40K. \newline In a recent NMR work\cite
{Barbara99}, we have shown that some of the $^{133}{\rm Cs}$
sites remain unaffected by the onset of the spin-ordering in the
low temperature state, magnetic and nonmagnetic domains being
spatially distributed. At the temperature of 13.8K the occurrence
of a charge redistribution and a concomitant decrease of the
local electronic susceptibility inside these nonmagnetic domains
have been observed \cite{Barbara99}. In agreement with this
latter result, detailed analysis of the EPR linewidth at ambient
pressure also suggest that two distinct magnetic environments
coexist in the low temperature state of ${\rm RbC}_{60}$ and ${\rm
CsC}_{60}$ polymers \cite{Atsarkin97,Coulon00} and insofar as a
charge redistribution occurs in the nonmagnetic domains at
13.8K\cite{Barbara99}, the spontaneous thermal contraction
recently observed at 14K by X-Ray diffraction in ${\rm
CsC}_{60}$\cite{Rouziere00} strongly supports the fact that these
inhomogeneities are intrinsic.\newline In this manuscript, we
give experimental evidence showing that the ``conducting'' state
of the ${\rm CsC}_{60}$ polymerized phase cannot be understood
within the framework of an electronic band conductor as claimed
earlier\cite{Chauvet94,Bommeli95,Brouet96,Janossy97,Bennati98,Erwin95}
. We first report the temperature dependence of the spin lattice
relaxation rate $(T_{1})^{-1}$ for both $^{13}{\rm C}$ and
$^{133}{\rm Cs}$ nuclei at different pressures up to 9 kbar,
indicating that in the temperature domain above 80K two
different mechanisms govern the relaxation of $^{13}{\rm C}$ and $^{133}{\rm %
Cs}$ nuclei respectively. As far as $^{133}{\rm Cs}$ is concerned,
$^{133}(T_{1})^{-1}$ decreases linearly down to about 80K though
remaining pressure independent up to 9kbar. This behavior is in
sharp contrast with the $^{13}{\rm C}$ nuclei for which
$^{13}(T_{1})^{-1}$ strongly decreases under pressure up to 9kbar
while remaining almost temperature independent. The difference
between $^{133}{\rm Cs}$ and $^{13}{\rm C}$ nuclei exists {\em
independently} of the nature of the ground state of the system.
More insight into these peculiar
properties is then obtained using quadrupolar echo experiments performed on the $%
^{133}{\rm Cs}$ nucleus which enable us to analyze with great
accuracy the temperature dependence of the NMR spectrum at 1bar.
We show that the NMR spectrum of the two phases (magnetic and
nonmagnetic) is motional narrowed
above 100K because of the fast motion of the local environment around the $%
^{133}{\rm Cs}$ sites. The evolution of the lineshape with
temperature reveals that the static coexistence of two different
$^{133}{\rm Cs}$ sites below 15K arises from a gradual freezing of
these fluctuations in the local environment.

\section{EXPERIMENTAL DETAILS}
 The measurements have been conducted on two powdered
samples with entirely consistent results, one of them (10\%)
$^{13}{\rm {C}}$ enriched. The pressure set up is a homemade
double-stage copper-beryllium cell using fluor-inert as the
pressure medium. This enables us to correct for each temperature
the loss of pressure within the sample chamber due to the gradual
freezing of the fluor-inert.\newline The spin-lattice relaxation
were measured by monitoring the recovery of the magnetization
after saturation with a series of $\pi/2$ pulses. The recovery
curve is exponential for $^{133}{\rm Cs}$ and $^{13}{\rm C}$
\cite{comment1} at room temperature. At ambient pressure, the
recovery curve gradually becomes bi-exponential for both nuclei
below 40K. A large distribution of short relaxation rates is
observed giving raise to a recovery curve of the following shape
$1-e^{{(-t/T_{1})}^\beta}$ with a value of $\beta$ of the order
of 0.5 at the lowest temperature investigated i.e 4K. At 5kbar
the recovery curve is for $^{133}{\rm Cs}$ exponential down to
4K. Not so for $^{13}{\rm C}$  since a nonexponential recovery is
observed below 20K. Different fit procedures did not help us to
determine without ambiguity the shape of the recovery but no
significant change were observed on the qualitative temperature
dependence of $^{13}(T_{1})^{-1}$. The relaxation rates
$^{13}(T_{1})^{-1}$ shown on Fig.1b at 5kbar are therefore
deduced below 20K from a fit of the recovery curve assuming it to
be exponential as above 20K. At 9kbar, the recovery curve is
exponential for $^{133}{\rm Cs}$ and $^{13}{\rm C}$  in the all
temperature range investigated. \newline Finally we should point
out that despite the presence of a static quadrupole
splitting\cite{comment2} for the NMR line of $^{133}{\rm Cs}$ in
the orthorhombic phase, the smallness of the quadrupole frequency
which is of the order of 5kHz enables us to saturate all the
transitions at once. Therefore the nuclear levels are initially
equally populated establishing a well-defined spin temperature
(equal to infinite). In that case no deviation from an exponential
behavior is expected for the relaxation of the
magnetization\cite{Suter98} which
perfect exponential recovery at room temperature is a proof of the homogeneity of the samples.

\section{$^{13}\rm C$ and $^{133}\rm Cs$-NMR UNDER PRESSURE}
We report on Fig.1a and Fig.1b the temperature dependence of the
relaxation rate for $^{133}{\rm Cs}$ and $^{13}{\rm C}$ nuclei at
ambient pressure, 5kbar and 9kbar. The large enhancement of
$^{133}(T_{1})^{-1}$ and $^{13}(T_{1})^{-1}$ below 40K is due to
a slowing down of magnetic fluctuations which is completely
suppressed at 5kbar. At this pressure, both $^{133}(T_{1})^{-1}$ and $%
^{13}(T_{1})^{-1}$ decrease exponentially below 20K revealing the
opening of a spin-gap at $T_{C}\approx 20K$, the ground state
being homogeneous and nonmagnetic. The effect of an applied
pressure on this long range order has been carefully investigated
by $^{133}{\rm Cs}$-NMR. The temperature dependence of
$^{133}(TT_{1})^{-1}$ at 5, 5.5, 5.7 and 9kbar is shown in
Fig.2.The well-defined
instability at 5kbar gives rise to a sharp peak on $^{133}(TT_{1})^{-1}$ at $%
T_{C}$. Quite remarkably, a slight increase of the applied
pressure strongly
reduces the amplitude of the spin-gap, without any significant change in $%
T_{C}$ itself (as given by the position of the
$^{133}(TT_{1})^{-1}$ peak, see on Fig.2). However, a smooth
decrease of the temperature ${\rm T_{Mag}}$ at which the slowing
down of magnetic fluctuations occurs has been observed by EPR
experiments under pressure up to 4kbar\cite{Forro96}. This fact is
also evident from the temperature dependence of the linewidth of
the $^{133}{\rm Cs}$ NMR line shown at different pressures on
Fig.3. Thus, as the pressure increases ${\rm T_{Mag}}$ drops
continuously along a transition line which does not exist for the
case of the spin-singlet ground state. Henceforth we can infer
that the sharp suppression of the spin-gap below 20K which in
turn gives rise to a metallic state, is not due to continuous
changes in the magnitude of the electronic interactions but may
reflect some structural changes above 5kbar as suggested by DC
conductivity measurements performed under
pressure\cite{Khazeni97,Zhou00}.

\noindent Another striking feature in the response of the polymerized phase $%
{\rm CsC}_{60}$ to high pressure appears at glance in Fig.1a and
Fig.1b. Indeed, a clear distinction has to be made between the two
temperature
domains 4.2-80K and 80K-300K. Below 80K, $^{133}(T_{1})^{-1}$ and $%
^{13}(T_{1})^{-1}$ exhibit a similar pressure and temperature
dependence.
This is, however, no longer true above 80K, where $^{133}(T_{1})^{-1}$ and $%
^{13}(T_{1})^{-1}$ behave in complete different ways. In
particular, we can see in Fig.1a that above 80K,
$^{133}(T_{1})^{-1}$ shows no pressure dependence up to 9kbar
unlike $^{13}(T_{1})^{-1}$, which is shown on Fig.1b. Within the
first five kilobars, the relaxation of $^{13}{\rm C}$ nuclei is
strongly affected by pressure in two manners:(i) an overall
depression is observed under pressure following the depression of
the uniform spin susceptibility ($\chi $) measured by EPR
\cite{Forro96} which drops at a rate of about 10\% per kbar, (ii)
a weakly temperature dependent contribution to $^{13}(T_{1})^{-1}$
(20\% decrease from 300 to 40K) is suppressed at 5 kbar.\newline Broadly
speaking, the spin-lattice relaxation rate for a given nuclei and
the static electronic spin susceptibility are linked together by
the following relation : $(T_{1}T)^{-1}\propto
\sum_{q}|A(\vec{q})|^{2}\chi{_{\perp}} ^{^{\prime \prime
}}(\vec{q})$ where
$A(\vec{q})=\sum_{i}A_{i}e^{i\vec{q}.\vec{r}_{i}}$ is the form
factor of the  hyperfine interaction between a given nuclei and
the electronic spins located at the neighboring sites. Unlike
$^{133}{\rm Cs}$ which environment is octahedral, there is no
particular symmetry for $^{13}{\rm C}$ sites. If both nuclei are
coupled to the same electronic spins then, that
$^{13}(T_{1})^{-1}$ and $^{133}(T_{1})^{-1}$ display a different
pressure and temperature dependence above 80K,  might be
attributed to the presence of  a spatially dependent electronic
spin susceptibility which dominates the relaxation of $^{13}{\rm
C}$ nuclei. However, as previously shown for ${\rm
RbC}_{60}$ \cite{Senzier96},
  the decrease of $^{13}(T_{1})^{-1}$ follows  the decrease of
  the uniform spin  susceptibility deduced from EPR\cite{Forro96} within at least the first five
  kilobars. This reveals that in the low pressure regime, the dominant
contribution to the relaxation of $^{13}\rm C$ above 80K is due to enhanced
magnetic fluctuations at the wave vector $\vec{q}=0$ and therefore,
the differences described above between $^{13}\rm C$ and
$^{133}\rm Cs$ cannot be ascribed to the form factor of the alkali
site in the polymerized phase. \newline As it is, one can draw the
following conclusions. First, the absence of pressure dependence
observed for $^{133}(T_{1})^{-1}$ above 80K shows that the
dominant contribution to the fluctuating field at $^{133}{\rm
Cs}$ site in this temperature range is unrelated to the
electronic spins involved in the relaxation of $^{13}{\rm C}$
nuclei. Secondly, the fact that above 80K, $^{13}(T_{1})^{-1}$ is
weakly temperature dependent at ambient pressure and constant at
5kbar suggests that the electronic spins are localized. This
latter conclusion is in agreement with the calculated band
structure of the polymer $({\rm C}_{60}^{-})^{n}$\cite{Stafstrom95}
which displays a
 dispersionless 1D half-filled band at the Fermi level but in apparent contradiction
 with
  transport measurements\cite{Khazeni97,Zhou00} performed in the similar compound ${\rm
  RbC}_{60}$.\newline
 One can therefore conclude that a model based on a single electron specie is inadequate for
  describing the electronic properties of the polymerized phases ${\rm
RbC}_{60}$ and ${\rm CsC}_{60}$.  \newline
In a previous work\cite{Barbara99}, we have shown that the use of quadrupolar
spin echoes of $^{133}{\rm Cs}$ nuclei enables to reveal the
presence of nonmagnetic domains within a magnetic background.
However, whether this inhomogeneous state results from the
existence of static structural defects along the chains or is
purely electronically driven e.g. as proposed for underdoped
cuprates\cite{Emery} and spin-ladders compounds\cite{Scalapino},
remained an open question. In what follows, we address this
problem again with the aid of quadrupolar spin echoes in order to
determine how the inhomogeneous state at low temperature arises
from the high temperature one.

\section{$^{133}\rm Cs$-NMR AT AMBIENT PRESSURE}

In a similar way than in the reference\cite{Barbara99}, the spin echoes of $^{133}{\rm Cs}$ have been obtained after a ($%
\pi /2-\tau -\pi /8$) in-phase RF pulse sequence\cite{comment3},
maintaining fixed the echo delay $\tau $ at 40$\mu s$. Half of
the spin-echo at 3$\tau $ is then Fourier transformed. This
procedure gives rise to a spectrum containing two lines
$5/2\rightarrow 3/2$ and $-3/2\rightarrow -5/2$ split
by an amount $4\nu _{Q}$, where $\nu _{Q}$ is the quadrupole frequency of $%
^{133}{\rm Cs}$ nuclei in the polymerized phase
\cite{Barbara99,Barbara}. \noindent The evolution of the
$^{133}{\rm Cs}$ spectrum is displayed on Fig.4 at different
temperatures between 100 and 4.2K. The expected doublet spectrum
corresponding to a single $^{133}{\rm Cs}$ site is observed at
100K, but as $T$ approaches 40K, the shape becomes asymmetric and
a fine
structure gradually develops. At 25K, the coexistence of two different $%
^{133}{\rm Cs}$ sites is evident in Fig.4, with a frequency
difference in the local field of the order of $4\nu _{Q}$. This
means therefore that two distinct magnetic environments are
spatially distributed at this temperature. As the temperature is
further lowered, the situation with a single quadrupolar split is
recovered and thus only one $^{133}{\rm Cs}$ site contributes to
the spin echo signal below 15K. The amplitude of the spin echo
refocused at 3$\tau $ is proportional to $e^{-3\tau \gamma \Delta
H(T)}$ where $\Delta H(T)$ is the width of the local field
distribution due to the static electronic moments at a given
temperature~$T$. Considering two
distinct populations of $^{133}{\rm Cs}$ nuclei below 30K, $N_{m}$ and $%
N_{nm}$ which are coupled to the local field inhomogeneity $\Delta
H(T)$ and
located inside the nonmagnetic domains respectively, the total number of $%
^{133}{\rm Cs}$ sites contributing to the spin echo signal at
3$\tau $ can be expressed as:~$N(T)=N_{m}/(1+(3\tau \gamma \Delta
H(T))^{2})+N_{nm}$. If the condition $3\tau
\gamma \Delta H(T)\gg
1$ is fulfilled, only a fraction $N_{nm}$ of the nuclei contribute
to a spin echo at $3\tau $ since this experiment selects those
${\rm Cs}$ sites which are entirely decoupled from the onset of
local magnetism. Let $I(T)$ be the integrated intensity of the
Fourier transform performed on this spin echo. The temperature
dependence of $N(T)$
(equal to $I(T).T$) is reported on Fig.5. We observe that a majority of the $%
^{133}{\rm Cs}$ nuclei is gradually wiped out of the signal below
40K. A minimal value for $N_{nm}$ is reached at 15K and amounts to
about 10\% of the total number of nuclei at $40K$. However, the
estimated ratio between the two phases from the $^{13}\rm C$
spectrum suggests that approximately half of the $^{13}\rm C$ sites
do not see the magnetic moment distribution in the low temperature
state\cite{Brouet96}. We may solve this puzzle by considering that
the $^{13}\rm C$ spins probe the very local properties within each
$\rm C_{60}$ chains carrying the electronic spins whereas only
${^{133}\rm Cs}$ sites far from
any magnetic domain will contribute to the echo signal refocused at $3\tau $%
. This would mean that the boundary surface is large compare to
the domains
size suggesting that the phase separation sets on a microscopic scale.%
\newline
To gain insight into the driving force of this process more
attention must be paid to what happens above the spin ordering
temperature. In particular, we see on Fig.4 that the splitting of
the $^{133}{\rm Cs}$ spectrum displays a fine structure near 40K
although the NMR spectrum corresponding to two $^{133}{\rm Cs}$
sites is not yet resolved. This can be understood if we assume
that the local field of a $^{133}{\rm Cs}$ nucleus jumps randomly
from one value to the other in the ``conducting'' state. Indeed,
using only the difference between the resonance frequencies
$\delta \omega $ and the hopping time $\tau _{h}$, we can propose
the following scenario. At high temperature, $\delta \omega \tau
_{h}\ll 1$ and the spectrum is motional narrowed, which means that
only one doublet is visible. When the temperature
is lowered, the jump frequency ($1/\tau _{h}$) decreases and the condition $%
\delta \omega \tau _{h}\approx 1$ becomes fulfilled with a fine
structure
developing in the NMR spectrum. Finally, when $\delta \omega \tau _{h}\gg 1$%
, the quadrupolar splitting of the two sites are well resolved,
{\it i.e}.
one for $^{133}{\rm Cs}$ sites in the magnetic domains and the other for $%
^{133}{\rm Cs}$ sites in the nonmagnetic ones. We simulate each of
the three cases and our simulations at fixed $\delta \omega $ are
shown in Fig.6 for different correlation times $\tau _{h}$ and
superimposed (dotted line) on the experimental spectra on Fig.4.
Clearly, the calculated spectra bear a strong resemblance with
the experimental ones displayed on Fig.3 between 100 and 30K. We
can thus infer the existence of a thermally activated change in
the local environment of $^{133}{\rm Cs}$ sites which may become
the dominant contribution to $^{133}(T_{1})^{-1}$ when the
frequency $1/\tau
_{h} $ is of the order of the Larmor frequency (43 MHz) of the $^{133}{\rm Cs%
}$ nuclei.  Therefore, from the results exposed in this section we
can conclude to the existence of another degree of freedom aside
from the  fluctuations of the electronic spins located on the
${\rm C}_{60}$ molecules, and possibly related to spontaneous
local structural changes in the polymerized phase.

\section{DISCUSSION}
 As emphasized above, one of the difficulty aroused by
our work is to bring together the conducting nature of the
polymerized phase established by optical and transport
measurements\cite{Bommeli95,Khazeni97,Zhou00} with the pressure
and temperature dependence of $^{13}(T_{1})^{-1}$ which strongly
suggest that electrons are localized. It therefore turns out
natural to question ourselves about the possible relationship
between the local structural change around  $^{133}{\rm Cs}$
nuclei and the presence of charge degrees of freedom like
polarons in the polymerized phase. On the basis of the above NMR
results and anticipating results described further on, we suppose
that the mobility of a charge carrier in the polymerized phase
mainly depends upon the occurrence of a local structural
distortion in its vicinity. From a point of view which is somewhat
naive, one may consider that at thermal equilibrium, the charge
carriers diffuse through the lattice under the action of a random
force $F(t)$ which takes on only two discrete values $\pm f_{0}$.
For our particular purpose, the relevant physical quantity to be
consider is the spectral density $F(\omega)$ defined as the
Fourier transform of the correlation function $\langle
F(t)F(t+\tau)\rangle$, the brackets indicating an ensemble
average. In our case, $\langle F(t)F(t+\tau)\rangle$ can be
assumed to be of the form\cite{Slichter}: $f_{0}^{2}
e^{-|t|/{\tau_{h}}}$, which leads to the following
spectral density : $F(\omega)=\tau_{h}/(1+(\omega \tau_{h})^2)$.
Because any excited state of the charge carriers is to relax due
to the random force $F(t)$, the spectral density $F(\omega)$ will
lead to a strong frequency dependence in the response function of
the carriers to external oscillating fields. It is therefore of a
great interest to focus on AC resistivity
measurements\cite{Zhou00} performed at ambient pressure in both
${\rm KC}_{60}$ and ${\rm RbC}_{60}$. For ${\rm KC}_{60}$ which
does not exhibit a slowing down of spin fluctuations, AC and DC
resistivities display a similar temperature dependence. This is
however not true for ${\rm RbC}_{60}$  since a frequency
dependent peak is clearly observed on AC resistivity. The peak
shifts from 35K at 1.1kHz down to 25K at 43Hz, the order of
magnitude of these frequencies being in good agreement with the
value we deduced from our simulate spectra in the same
temperature range for ${\rm CsC}_{60}$ (c.f.Fig.6). The fact that
the electronic properties of ${\rm RbC}_{60}$ and ${\rm
CsC}_{60}$ display similar electronic and structural features as
opposed to ${\rm KC}_{60}$ allows us to extrapolate the results
obtained by Zhou {\it et al} for ${\rm RbC}_{60}$ to the case of
${\rm CsC}_{60}$.  Thus experiments show that in the
two polymerized phases ${\rm RbC}_{60}$ and ${\rm CsC}_{60}$, the
dissipation reaches a maximum when the hopping frequency of the
local environment of the alkali ion becomes equal to the AC
frequency. Such a coincidence can be hardly fortuitous and
suggests that the mobility of the charge carriers in the
polymerized phase is strongly coupled to the
environment of the alkali ion. In this context, it is worthwhile to
mention that polaron-like distortions such as ${\rm
C}_{60}^{-1-x}-{\rm C}_{60}^{-1+x}$ have been predicted to be
energetically favorable in the charged polymer $({\rm
C}_{60}^{-})^{n}$ which exhibits a tendency to undergo a charge
density wave transition \cite{Springborg95}.
 In that particular case, the conduction mechanism would be due to an
  intramolecular property of the polymer itself and that would
drastically change our expectations regarding the pressure effect
on the electronic properties of the polymerized phase. However,
on the sole basis of the NMR experiment above described we cannot
address the microscopic mechanism at the origin of the
spontaneous formation of polarons in the polymerized phase.
\newline In the light of the above considerations, it is
interesting to shortly reconsider  the pressure effect on the
spin-lattice relaxation rate $^{13}(T_{1})^{-1}$  of $^{13}{\rm
C}$ nuclei in the low pressure regime. As mentioned above,
$^{13}(T_{1})^{-1}$ shows at room temperature a similar pressure
decrease than the electronic spin susceptibility deduced from EPR
\cite{Senzier96} which suggests that magnetic fluctuations
at the wave vector $\vec{q}=0$ dominate $^{13}(T_{1})^{-1}$ at
ambient pressure. One possible explanation for the origin of
these enhanced uniform fluctuations might be that polarons acting
as local defects, induce disorder in the AF exchange coupling $J$
along the chain leading to the formation of spin
clusters\cite{Theodorou77}. It was indeed shown
theoretically\cite{Theodorou77} that the low energy magnetic
fluctuations (i.e when $T\ll J$) of a disordered AF spins chain
are merely governed by clusters with an odd number of spins, each
one acting as a nearly free localized (1/2) spin. In such a case
the {\em reversible} suppression  at 5kbar of a weak temperature
dependent term in $^{13}(T_{1})^{-1}$ could be ascribed to the
suppression with applied pressure of disorder in the magnetic
coupling along the chain which presence would be henceforth
closely related to the slowing down of spin fluctuations in the
low temperature state. Much more experimental inputs are however
required to go beyond this statement.
\newline As it is, the phase
 separation occurring in the low temperature state at ambient pressure appears
 to be the logical outcome of the twofold nature of the polymerized phase ${\rm CsC}_{60}$ that is :
  mobile polarons spontaneously form aside from localized electrons and compete with a 3D magnetic order imposed by the transverse
 dipolar coupling between the chains. Note that the presence of nonmagnetic domains  is in itself a strong hint that polarons are not randomly
spatially distributed within the magnetic background but may form
collective structures developing a long range order below 14K as
suggested by NMR\cite{Barbara99} and X-ray
experiments\cite{Rouziere00}.

\section{Conclusion}
The work described in this manuscript deals with the electronic
 properties of the polymerized phase ${\rm CsC}_{60}$ extensively studied by NMR of $^{13}{\rm C}$
  and $^{133}{\rm Cs}$ nuclei. The salient result is that the electronic properties of the
 polymerized phase ${\rm CsC}_{60}$
involve two  degrees of freedom : one related to localized spins,
the other related to mobile charges which mobility is strongly
entangled to the local environment of the Cs ion.  The
polymerized phase ${\rm CsC}_{60}$ is therefore dynamically
inhomogeneous and as shown by NMR under pressure, this
feature persists up to 9kbar. \newline At ambient pressure static
inhomogeneities gradually develop below 40K concomitantly with a
slowing down of spin fluctuations.  At 5kbar, the polymerized
phase ${\rm CsC}_{60}$ undergoes  a nonmagnetic transition at
$T_{c}$ equal to 20K. The ground state is homogeneous and a spin
gap opened below 20K. Finally, a dramatic decrease of the
amplitude of the spin gap is observed above 5kbar without any
significant decrease of $T_{c}$. The presence of magnetism therefore
appears to be closely related to the occurrence of
static inhomogeneities. How does the applied pressure suppress these inhomogeneities and
stabilize a homogeneous nonmagnetic ground state? That cannot be
addressed by the present work but remains an important issue to
be solved.

\section{ACKNOWLEDGMENT}
 It is a pleasure to thank F. Rachdi for the $ ^{13}\rm
C$ enriched ${\rm C}_{60}$.  We are also very grateful to C.
Berthier, S. Brasovski, P. Carretta, P. Sotta and P. Wzietek
for illuminating discussions and to J. P. Cromi\`{e}res and M.
Nardone for technical assistance.\newline
One of the authors (L.F) is grateful for the support of the Swiss National Science Foundation.

\clearpage

\begin{figure}[h]
  \centering
  \includegraphics[width=0.48\linewidth]{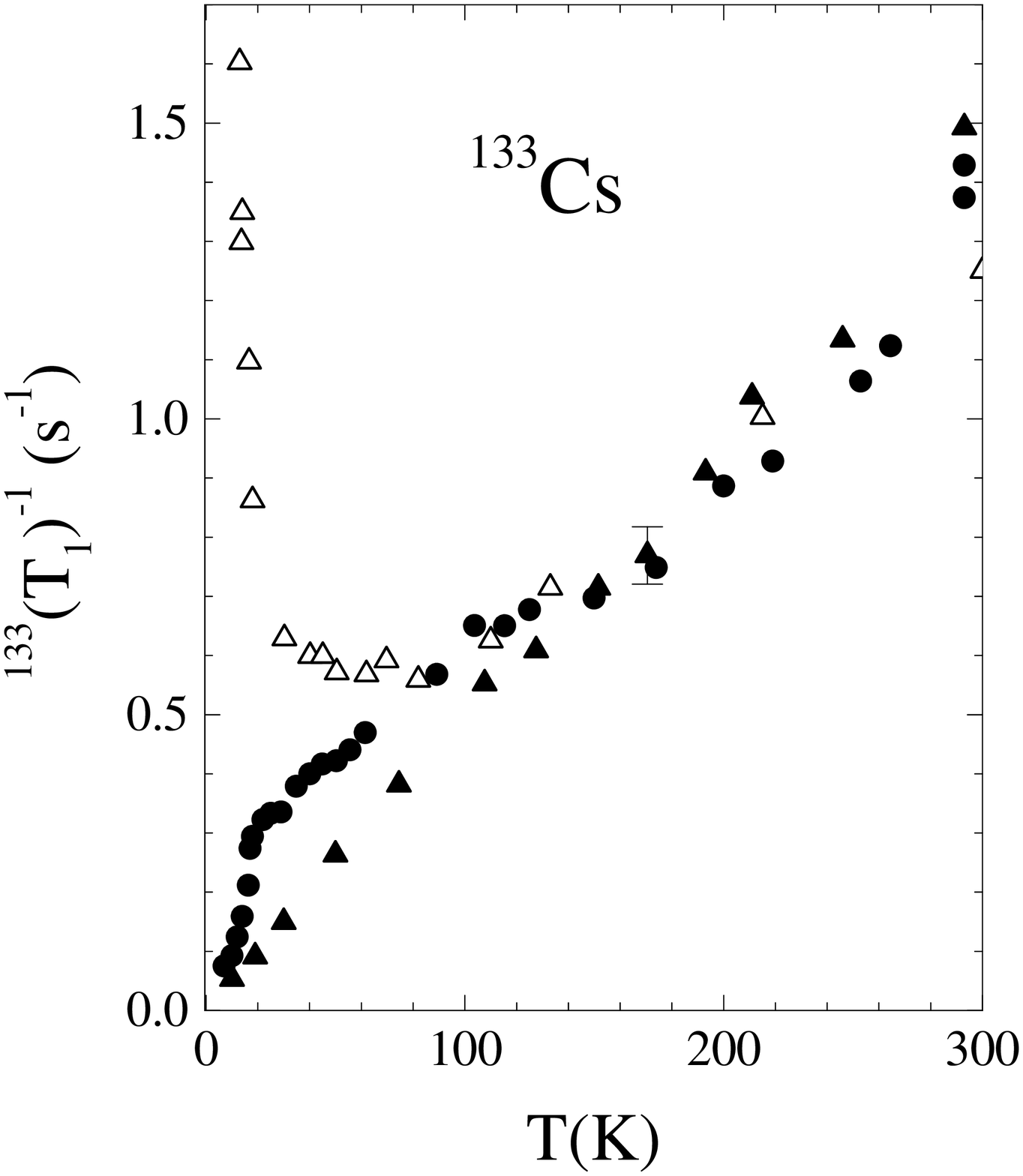}
  \includegraphics[width=0.48\linewidth]{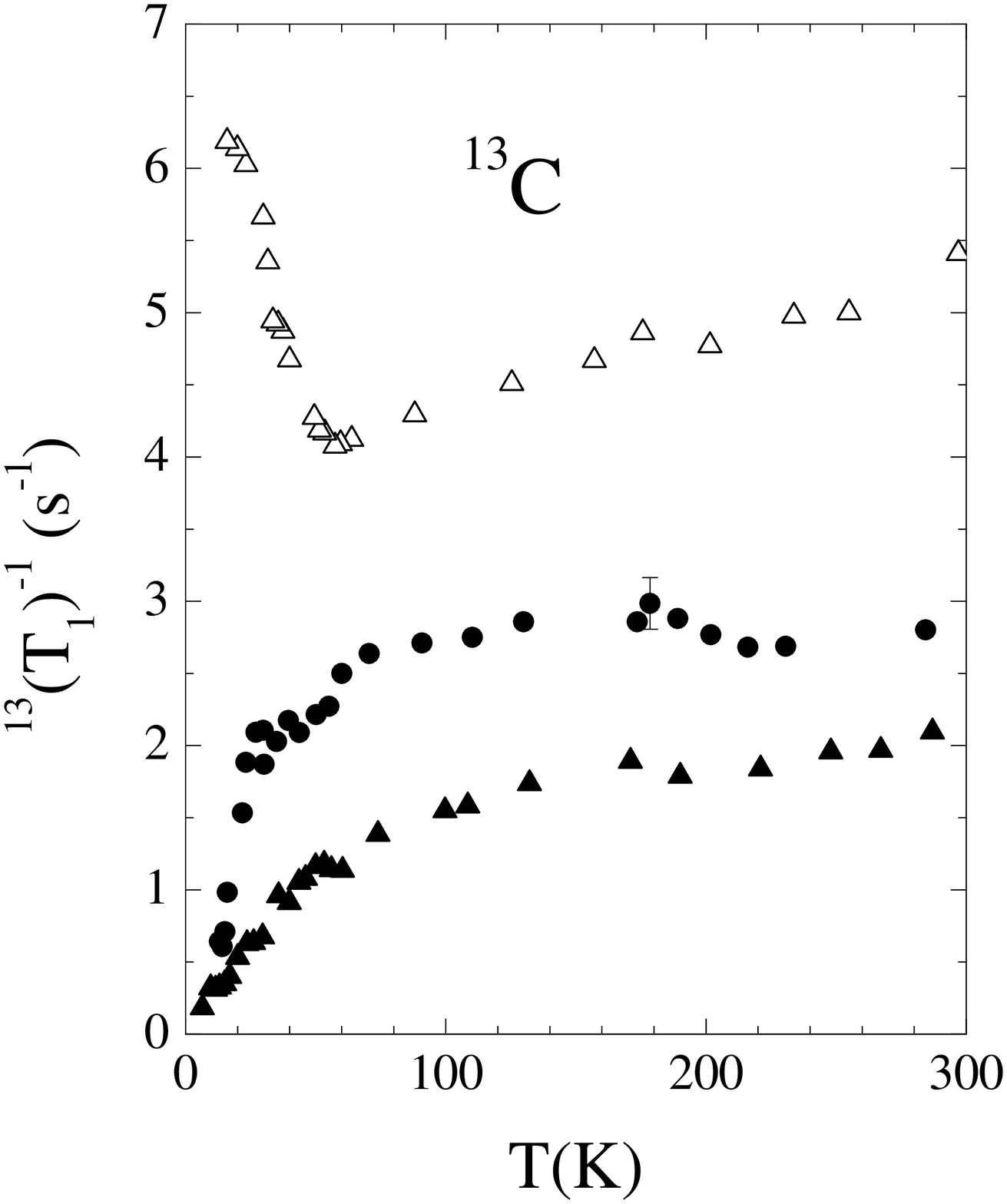}
  \caption{Temperature dependence of $T_{1}^{-1}$ for (a) $^{133}{\rm
     Cs}$ (at 8 Tesla) and (b) $^{13}{\rm C}$ (at 9 Tesla), at 1bar
     (empty triangles), 5kbar (black circles) and 9kbar (black
     triangles).  At 1bar and below 40K, the magnetization recovery
     curves are bi-exponential for both $^{133}{\rm Cs}$ and $^{13}{\rm
     C}$ but only the rapid component is reported versus temperature.}
\end{figure}

\begin{figure}[h]
  \centering
  \includegraphics[width=\linewidth]{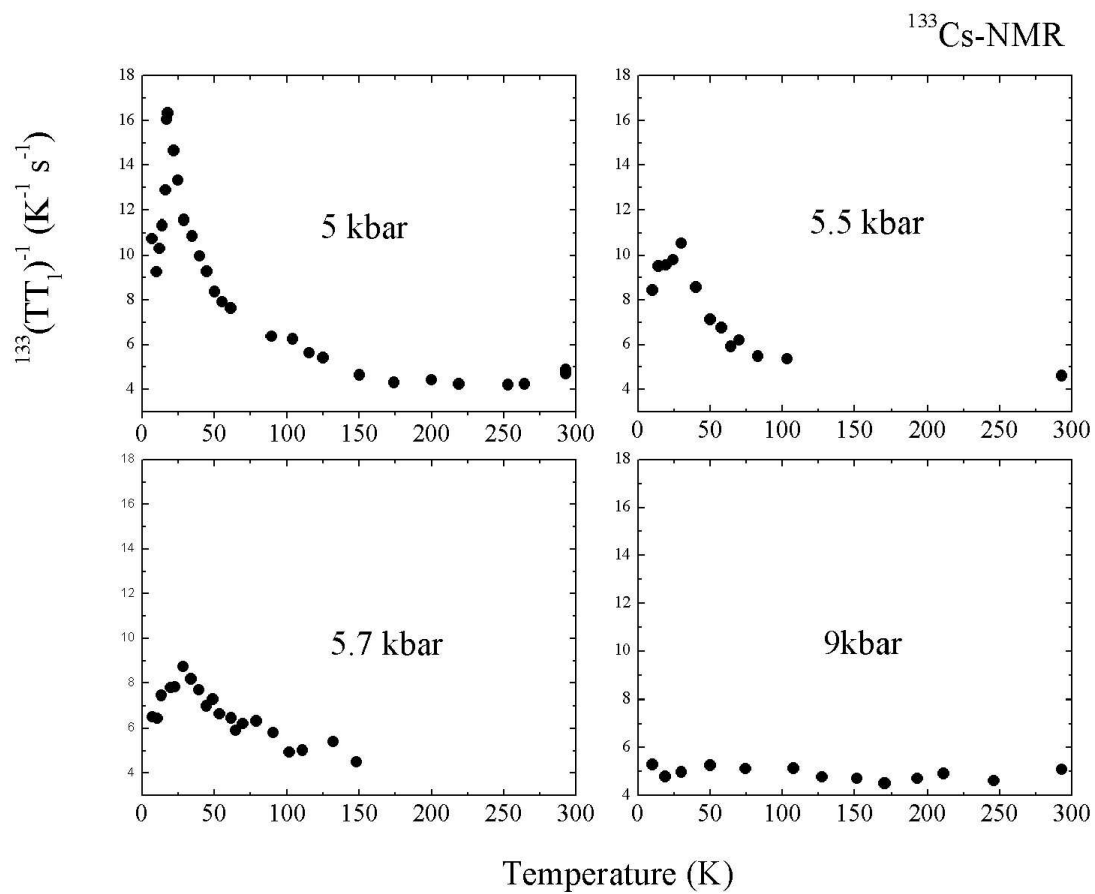}
  \caption{Temperature dependence of  $^{133}(TT_{1})^{-1}$  at 5kbar, 5.5kbar, 5.7 and 9kbar.}
\end{figure}

\begin{figure}[h]
  \centering
  \includegraphics[width=\linewidth]{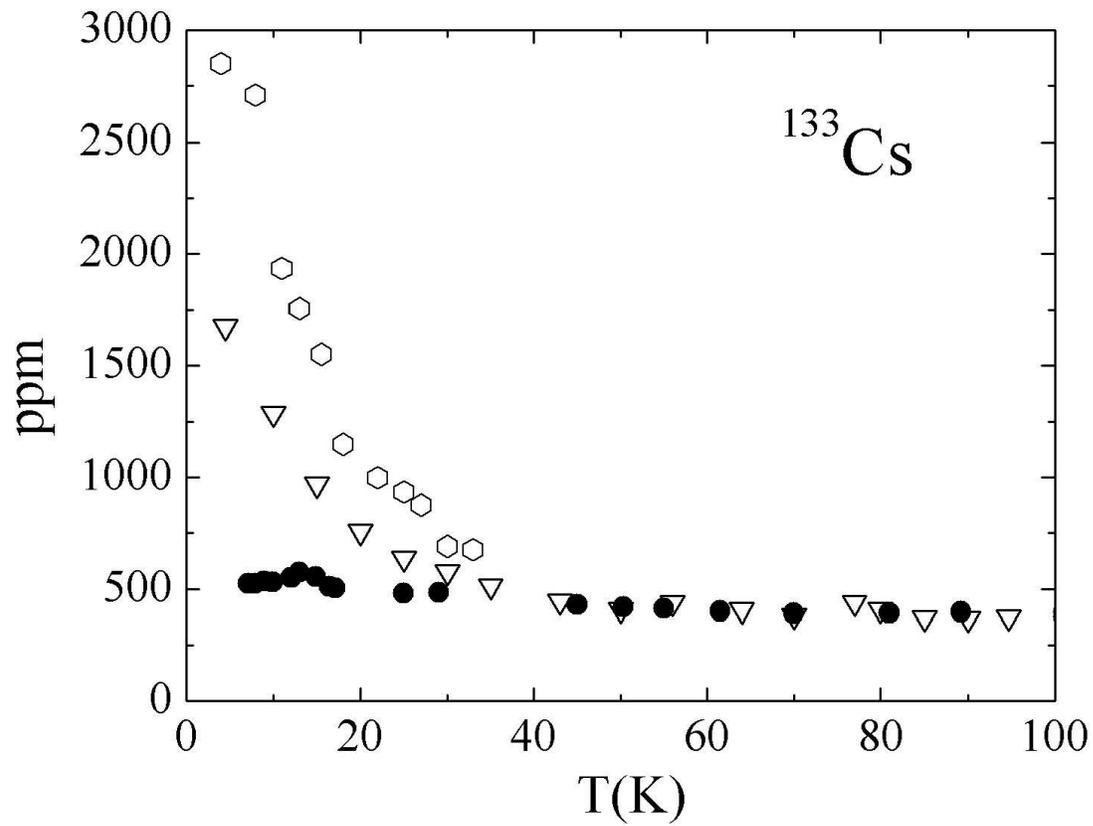}
  \caption{Temperature dependence of the linewidth of the $^{133}{\rm
Cs}$-NMR line at 1bar from ref~\protect\cite{Brouet96} (empty hexagons), at
3kbar (present study, empty triangles) and at 5kbar (present study,
full circles).}
\end{figure}

\begin{figure}[h]
  \centering
  \includegraphics[width=0.9\linewidth]{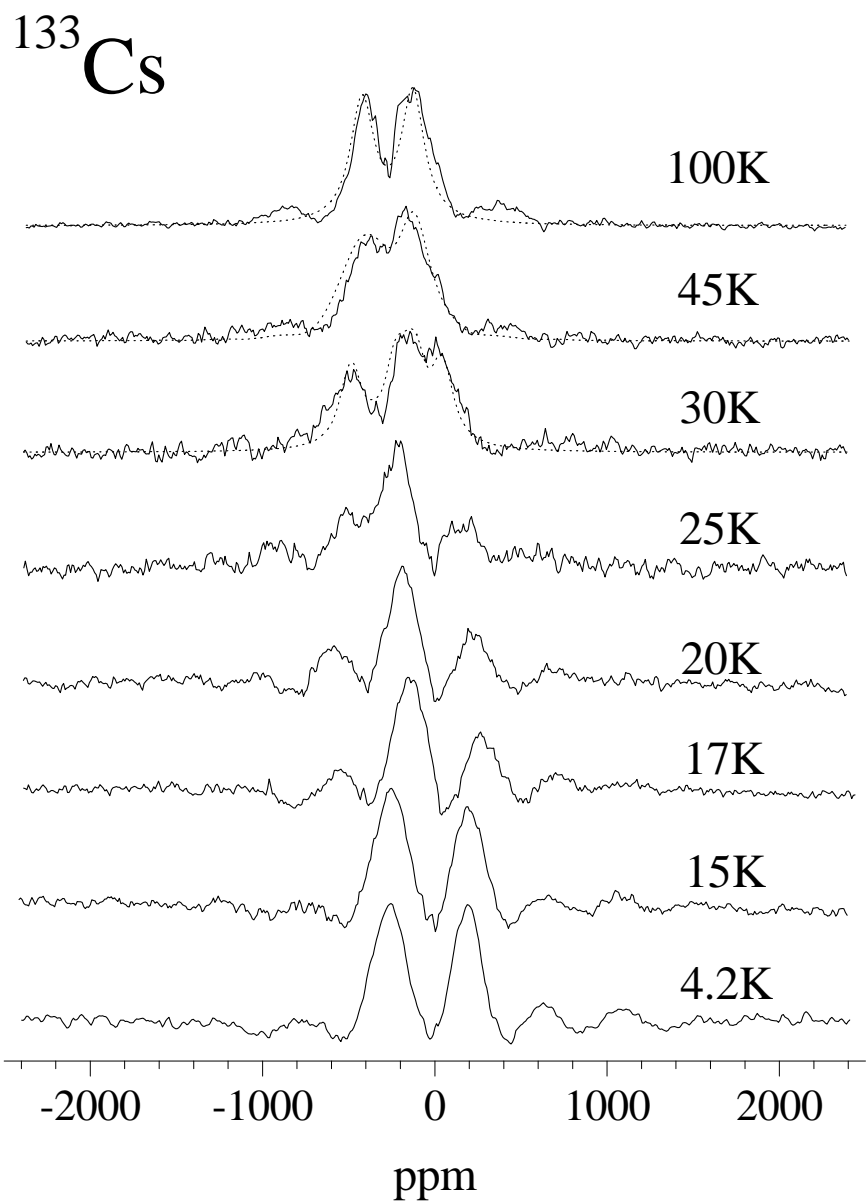}
  \caption{Evolution of the lineshape of the $^{133}{\rm Cs}$
quadrupolar normalized splitting from 100K down to 4.2K. Because
we take the Fourier transform of half of the spin echo at $3\tau$, the
other spin echoes induce distortion of the base line.}
\end{figure}

\begin{figure}[h]
  \centering
  \includegraphics[width=\linewidth]{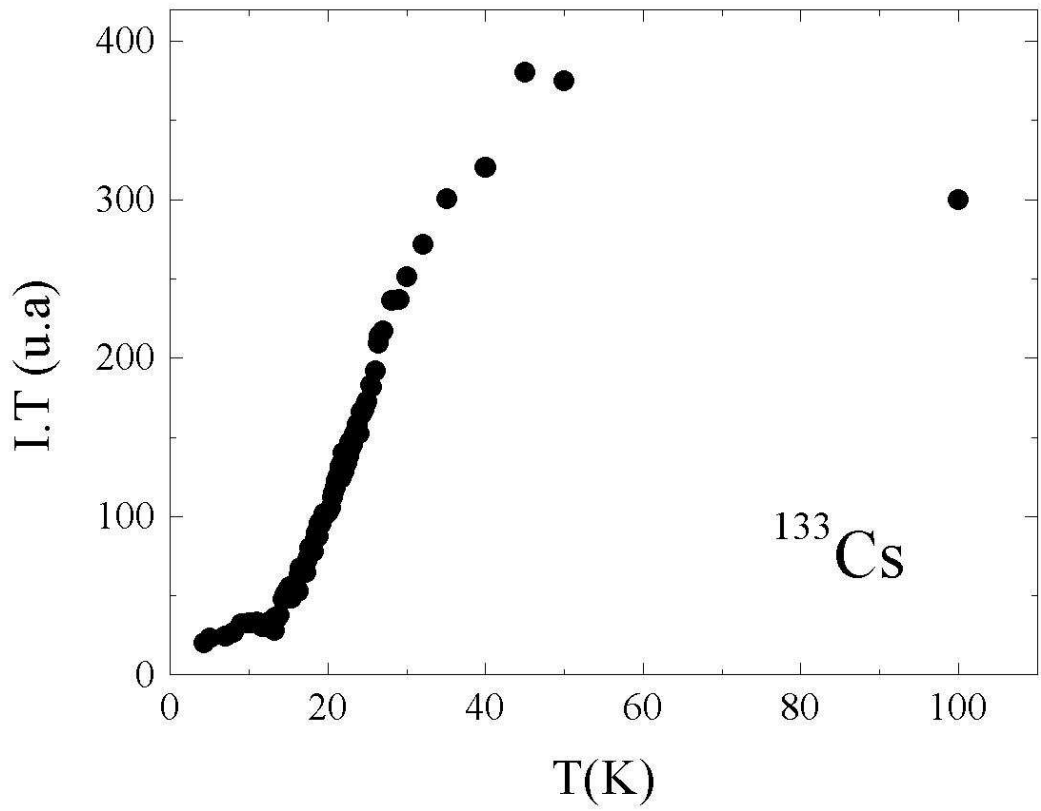}
  \caption{Temperature dependence of $N(T)=I(T).T$ where $I(T)$ is the
integrated intensity of the Fourier transform of half of the spin echo
refocused at $3\tau$.}
\end{figure}

\begin{figure}
  \centering
  \includegraphics[width=\linewidth]{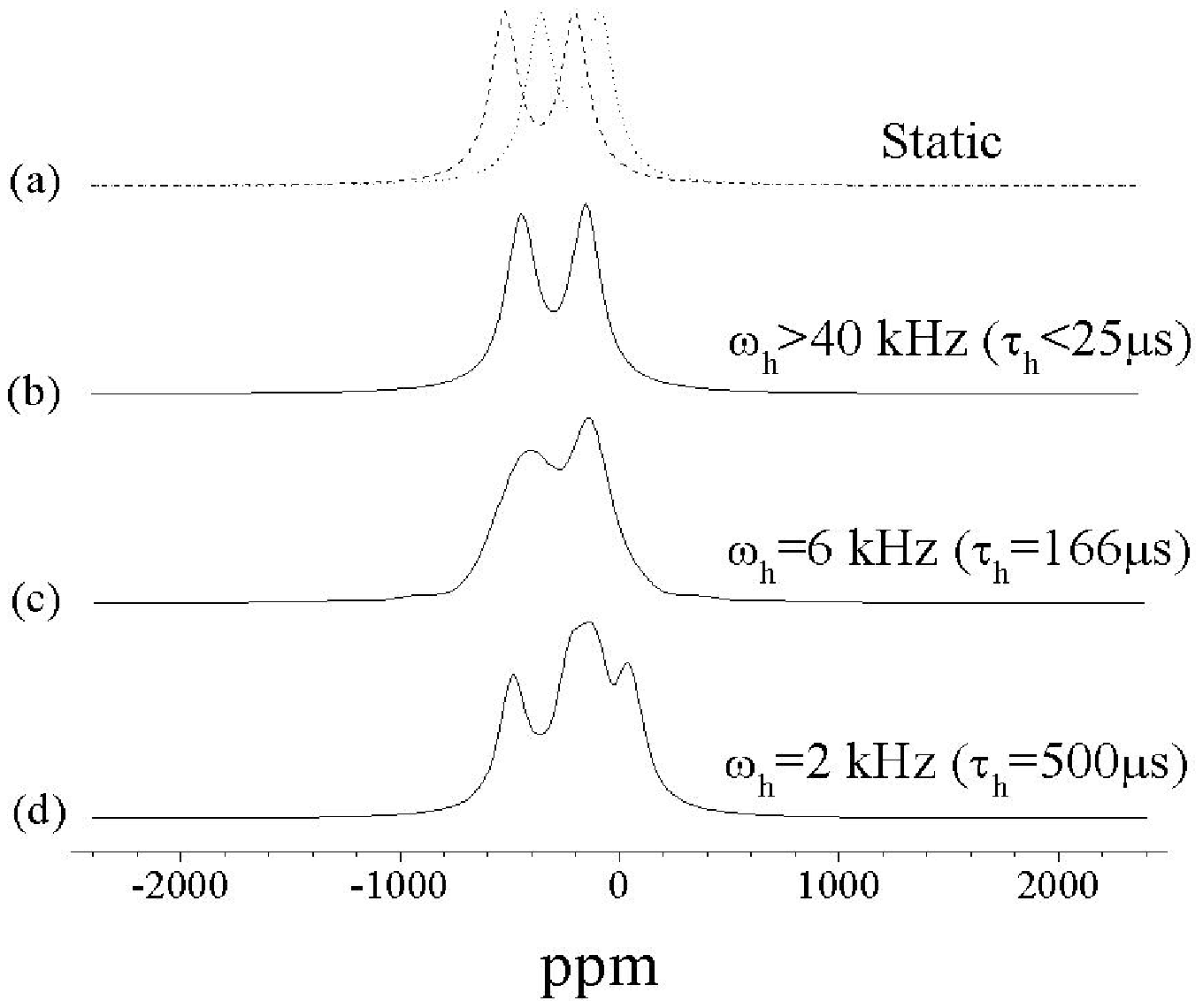}
  \caption{Simulation of the shape of the $^{133}{\rm Cs}$ quadrupolar
spectrum.(a) In the static case : for each of the two
configurations, both the quadrupole frequency $\nu_{Q}$ and the
frequency shift $K$ compare to the Larmor frequency are
determined from the spectrum at 100K {\it i.e.} in the fast motion
limit.This gives the respective values : $\nu_{Q}$=4.17kHz and
$K$=$-$350ppm (dashed line), $\nu_{Q}$=3.57kHz,
$K$=$-$209ppm (dotted line).(b) In the fast motion limit : $\delta \omega \tau_h \ll 1$%
.(c) In the slow motion limit : $\delta \omega \tau_h \approx 1$.(d)
For a {\em quasistatic} distribution : $\delta \omega \tau_h \gg 1$.
}
\end{figure}


\begin{references}
\bibitem{comment1}  We can always fit the recovery curve with the expression
$M(t)=M_{eq}(1-e^{(-t/T_{1})^{\beta}})$, $M_{eq}$ being the
equilibrium magnetization and $\beta$ a dimensionless coefficient
comprised between 0.5 and 1. At room temperature and at any
pressure investigated $\beta$ is equal to 1 for $^{133}$Cs
whereas a substantial deviation from an exponential recovery is
observed i.e $\beta \approx 0.72$ for $^{13}$C. Because of the
possible presence of a small amount of pure amorphous C$_{60}$,
such deviation might be extrinsic. Therefore, we talk about
"exponential recovery" for $^{13}$C as opposed to the dramatic
changes in the recovery curve observed at low temperature and
which occurrence is unambiguously related to some intrinsic
properties of the polymerized phase.
\bibitem{comment2} The presence of a quadrupole splitting does not
imply a quadrupole relaxation but is consequent to the noncubic
symmetry of the local environment of $^{133}$Cs site. For a cubic symmetry
the three components of the electric field gradient
 are strictly equal to zero which implies no quadrupole
splitting. However, any fluctuations leading to a substantial
deviation from the cubic symmetry may lead to a quadrupole
relaxation depending upon the magnitude of the quadrupolar coupling compare to others.
For the particular case of $^{133}$Cs, the
possibility of a quadrupole contribution in the relaxation of the
magnetization can be easily ruled out because of the remarkably
small value of its quadrupole moment.
\bibitem{comment3} For a given nuclei, the amplitude of the spin echoes strongly depends upon
 the angle of the rotation applied by the second pulse to the magnetization  as well as the relative phase between
  the two pulses. For the in-phase pulse sequence ($\pi /2-\tau -\pi
  /8$) applied on $^{133}{\rm Cs}$ nuclei, the calculated\cite{Barbara} amplitude of the spin echo is maximal at $3\tau$
  and strongly reduced at $5\tau/2$ and $4\tau$.
\end{references}
\end{document}